\documentclass[11pt]{article}

\usepackage{latexsym,color,graphicx,amsmath,amssymb,amsthm,graphics}
\usepackage{subfig}

\usepackage{chngcntr}
\counterwithin{figure}{section}
\counterwithin{table}{section}

\def\R{{\mathbb R}}

\theoremstyle{definition}

\graphicspath{{./figures/}}

\begin{document}

\title{Delaunay Triangulations of Degenerate Point Sets\thanks{%
Work on this paper was supported by Grant 892/13 from the Israel Science Foundation, 
by Grant 2012/229 from the U.S.--Israel Binational Science Foundation, by the Israeli 
Centers of Research Excellence (I-CORE) program (Center No.~4/11), and by the Hermann 
Minkowski-MINERVA Center for Geometry at Tel-Aviv University. The paper summarizes the results of the M.Sc. thesis of the first author, supervised by the second author.
}}

\author{
Michael Khanimov\thanks{%
School of Computer Science, Tel Aviv University,
Tel Aviv 69978, Israel.
{\sl michael1307@gmail.com} }
\and
Micha Sharir\thanks{%
School of Computer Science, Tel Aviv University,
Tel Aviv 69978, Israel.
{\sl michas@post.tau.ac.il} }
}


\maketitle

\begin{abstract} 

The Delaunay triangulation (DT) is one of the most common and useful triangulations of point sets $P$ in the plane. DT is not unique when $P$ is degenerate, specifically when it contains quadruples of co-circular points. One way to achieve uniqueness is by applying a small (or infinitesimal) perturbation to $P$.

We consider a specific perturbation of such degenerate sets, in which the coordinates of each point are independently perturbed by normally distributed small quantities, and investigate the effect of such perturbations on the DT of the set. We focus on two special configurations, where (1) the points of $P$ form a uniform grid, and (2) the points of $P$ are vertices of a regular polygon.

We present interesting (and sometimes surprising) empirical findings and properties of the perturbed DTs for these cases, and give theoretical explanations to some of them.

\end{abstract}


\section{Introduction}

\begin{figure}[h]

\centering

\includegraphics[width=0.7\textwidth]{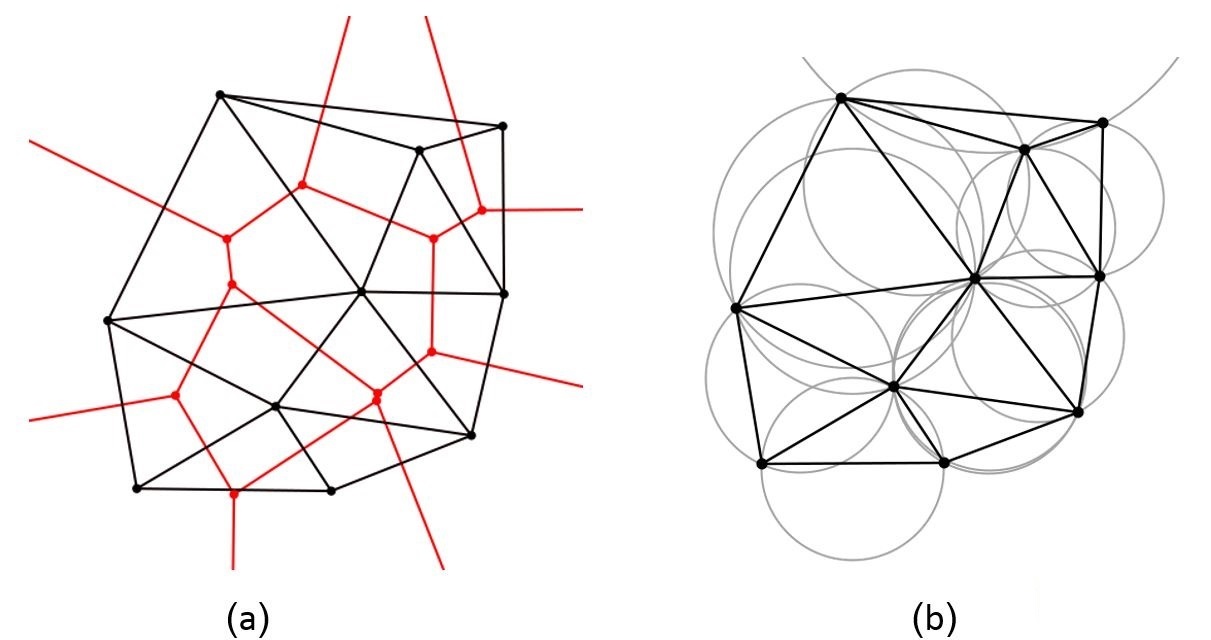}

\caption{\small (a) Delaunay triangulation and Voronoi diagram. (b) Delaunay triangulation with all the circumcircles of its triangles. The interior of each circumcircle contains no input points. The figures are taken from \cite{wiki}.}

\label{fig1.1}

\end{figure}

\subsection{Basic definitions}

\label{sec:intro_background}

\paragraph{Delaunay triangulation.} Let $P$ be a set of $n \geq 3$ distinct points in the plane. A triangulation of $P$ is called a 
\emph{Delaunay triangulation} (and denoted $DT(P)$) iff it satisfies \textbf{\emph{the empty circumcircle criterion}} that is defined as follows: the interior of the circumcircle of any triangle in the triangulation contains no point of $P$. See an example in Figure \ref{fig1.1}.

\paragraph{Degeneracy and general position.}In the context of Delaunay triangulations, $P$ is said to be \emph{degenerate} (or not in \emph{general position}) if either it contains four co-circular points, such that the circle through them contains no other point of $P$ in its interior, or it contains three collinear points on the boundary of its convex hull. $DT(P)$ is not unique when $P$ is degenerate.

\paragraph{The Incircle test \cite{GS1985}.} This is a simple and practical test that determines whether a point $D$ lies inside, outside, or on the circle
defined by three other distinct points $A$, $B$ and $C$ in the plane. Concretely, given that the points $A$, $B$, and $C$ define a CCW oriented triangle,
the predicate $Incircle(A,B,C,D)$ is defined to be true iff point $D$ lies inside the circle passing through $A$, $B$, $C$. Guibas and Stolfi \cite{GS1985}
show that the test $Incircle(A,B,C,D)$ is equivalent to
$$
\begin{vmatrix}
x_A & y_A & x_A^2 + y_A^2 & 1\\
x_B & y_B & x_B^2 + y_B^2 & 1\\
x_C & y_C & x_C^2 + y_C^2 & 1\\
x_D & y_D & x_D^2 + y_D^2 & 1\\
\end{vmatrix} > 0,
$$ where $x_p, y_p$ are the coordinates of point $p \in \{A,B,C,D\}$. The determinant is negative iff $D$ lies outside the circle, and equals zero iff $A$, $B$, $C$, and $D$ are cocircular.

\paragraph{The normal perturbation.} Given a set of points $P$ in the plane, a perturbation of $P$ is obtained by shifting each $p \in P$, independently at random, by small (or infinitesimal) quantities $dx, dy$ in the $x, y$ coordinates, respectively. We use a specific perturbation in which $dx, dy$ are distributed according to $0.001 \cdot N(0,1) \cdot d_{min}$, where $N(0,1)$ is the standard normal distribution (with mean 0 and standard deviation 1), and $d_{min}$ is the minimum distance between any pair of points of $P$. We refer to this perturbation as the \textbf{\emph{normal perturbation}}. We denote by $P'$ a (random) set of points that were obtained from $P$ after it passed the normal perturbation. From the continuity of the normal distribution, such a perturbation guarantees a general position with probability 1. So, with probability 1, $P'$ has a unique DT.

\subsection{Our work}

\paragraph{The goals.} We take one of the simplest ways to resolve degeneracy, that is, by a random small perturbation of the input of the sort considered above, and investigate its influence on the resulting Delaunay triangulation.

We chose to focus on two types of degenerate sets: a uniform grid of points and the vertex set of a regular polygon. The reason for these choices is that on one hand these sets are basic and simple to describe and analyze, and on the other hand they are clearly (very much) not in general position.

\paragraph{The results.} For the point sets we studied, we show that, after applying the normal perturbation on the input, the different Delaunay triangulations that can arise are in general far from being uniformly distributed.

A significant part of this work has been experimental, where we applied many random normal perturbations to $P$, and collected empirically the resulting DT distribution. In a few cases we were also able to explicitly (and accurately) calculate the DT distribution and, in this way, to validate our empirical findings. For several other cases, involving larger number of points, we just proved that their distribution is different from the uniform one, without specifying what it is exactly. And for the rest, we just had to do with our empirical findings.

Our main conclusion is that the use of the normal perturbation, as a method to avoid degeneracy, should be accompanied with the awareness that this method introduces a bias towards certain Delaunay triangulations over the others. Similar biases are likely to arise in other perturbation schemes of this sort. In addition, the analytic methods that we have used seem to be interesting in their own right, and we hope that they could find additional applications of this kind.

This research topic, in its general form, was suggested to us by Leo Guibas, and we thank him for that lead.


\section{Grids}

\begin{figure}[h]

\centering

\includegraphics[width=0.25\textwidth]{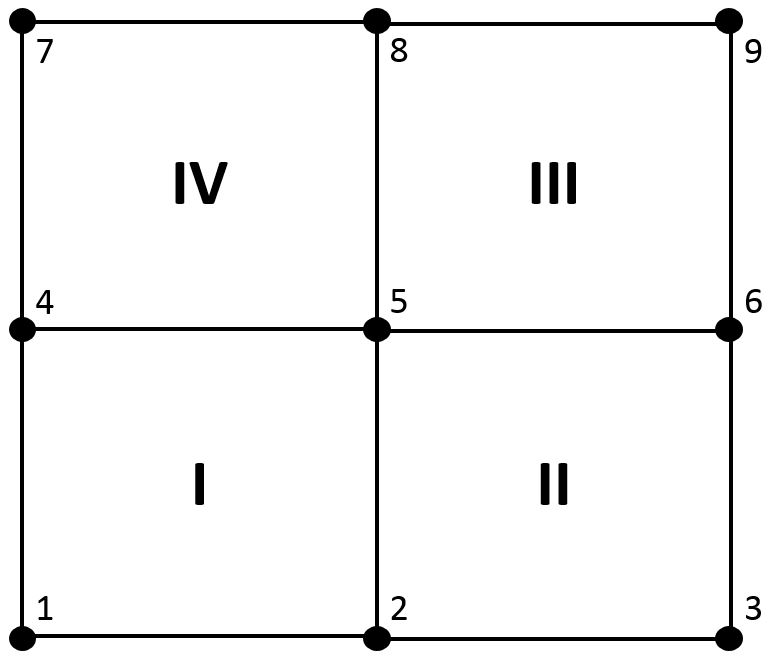}

\caption{\small A grid $P$ with $n=9$ and $m=2$.}

\label{fig2.1}

\end{figure}

Let $P$ be a uniform $(m+1) \times (m+1)$ grid of points in the plane, and put $n=(m+1)^2$. Without loss of generality, we assume that each cell of the grid is a unit square, that $(0,0) \in P$, and that $m \geq 1$ is a natural number. It is clear that in terms of the Delaunay triangulation, $P$ is (very) degenerate.

Let $C(P)$ be the graph that is obtained from $P$ by connecting every pair of horizontally or vertically consecutive points of $P$ by an edge. This ia a planar graph that has exactly $m^2$ bounded faces, each a square of size 1, and we denote them as the \emph{cells} of $C(P)$. From the empty circumcircle criterion, it holds that a triangulation of $P$ is a DT iff it is obtained from $C(P)$ by dividing each of its cells into two triangles by adding a diagonal (it is easy to see that all the edges of $C(P)$ are guaranteed to appear in every $DT(P)$). Thus, $P$ has exactly $2^{m^2}$ different Delaunay triangulations.

For each $T \in DT(P)$, our goal is to estimate the probability that $DT(P') = T$. Schrijvers \cite{S2014} showed that for every $T$ this probability is positive. The results reported below vary according to the value of $m$:

\subsection{Small Grids}

\paragraph{\boldmath$m=1$.} $C(P)$ is a single cell, so $DT(P')$ is one out of two possible triangulations. By symmetry, the probability of each of them to appear is exactly 0.5.

\paragraph{\boldmath$m=2$.} The probability of any $T \in DT(P)$ to appear equals to the probability that for each cell $c \in C(P)$ its diagonals in $T$ and in $DT(P')$ are equal. This happens iff, after the perturbation, four Incircle tests hold simultaneously, one test for each $c$, where each test is on the vertices of $c$ in $P'$.

For convenience, we label the grid points and cells as in Figure \ref{fig2.1}. For example, informally, in order that cell $I$ contains the diagonal `/' in $DT(P')$, in $P'$ point 4 should lie outside the circle that passes through the points $1,2,5$. That is, $Incircle(1,2,5,4) < 0$ should hold. On the other hand, in order that cell $I$ contains the diagonal `\textbackslash{}', $Incircle(1,2,5,4) > 0$ should hold. (Again, with probability 1 one of these inequalities must hold).

Let $x_i, y_i$ denote the shifts in the $x$- and $y$-coordinates of point $i$, as a result of the normal perturbation, and $\Delta(j)$ be the determinant that arises in the Incircle test that corresponds to cell $j \in \{I,II,III,IV\}$. After expanding the determinants of all the Incircle tests, we get that each of them is equal to the sum of products of the variables $x_i, y_i$ by themselves and by constants. (This sum does not have a constant term because, by the Incircle test definition, when all the variables $x_i, y_i$ are zero, $\Delta(j)$ is also zero.)

For the analysis, we are interested only in the sign of $\Delta(j)$ (not its magnitude). By the normal perturbation definition, all the $x_i, y_i$ are close 
enough to $0$, so the sign of $\Delta(j)$ is determined only by the sum of its first-order terms, as long as they are not too close to zero, as it is in our 
case (with probability almost 1). \footnote{In what follows, we ignore the small portion of the space of perturbations where the second-order terms may affect the Incircle tests, and assume that this never happens (or rather happens with probability 0). This leads to a small error in our estimates, which, as 
our experiments show, is indeed minuscule.}

After the evaluation of the determinants and the exclusion of terms of second or higher order, we get the following system:
\begin{align*}
\begin{cases}\tag{2.1}
sign(\Delta(I)) = sign(-x_1-y_1-x_2+y_2+x_4-y_4+x_5+y_5) \\
sign(\Delta(II)) = sign(-x_2-y_2-x_3+y_3+x_5-y_5+x_6+y_6) \\
sign(\Delta(III)) = sign(-x_5-y_5-x_6+y_6+x_8-y_8+x_9+y_9) \\
sign(\Delta(IV)) = sign(-x_4-y_4-x_5+y_5+x_7-y_7+x_8+y_8) . \\
\end{cases}
\end{align*}

For a given $T$ (i.e., a given choice of diagonals), the system (2.1) leads to four inequalities of the form ``$\Delta(j) \lessgtr 0$''. As stated above, the probability that $T = DT(P')$ equals to the probability that for each $\Delta(j)$, its sign in $P'$ corresponds to the diagonal direction of cell $j$ in $T$. In other words, $T = DT(P')$ iff in $P'$ the corresponding system of inequalities that is obtained from (2.1) holds.

\paragraph{Geometric approach.} We reduce this problem to a problem of calculating the volume of a certain object in 4-space. We represent each $P'$ by the
point $p' = (x_1,y_1,\ldots,x_9,y_9)$ in $\R^{18}$, i.e., the coordinates of $p'$ are the shifts $x_i, y_i$ of the points of $P$ in the perturbation. Each
inequality that is obtained from the system (2.1) defines a halfspace in $\R^{18}$ whose boundary passes through the origin. Let $S^*=S^*(T)$ be the intersection of these four halfspaces. The probability that $T = DT(P')$ equals to the probability of $p'$ to appear in $S^*$.

All the shifts $x_i, y_i$ are independent and normally distributed with the same distribution function. Thus $p'$ is distributed by the multivariate normal distribution in 18 dimensions. One of the properties of this kind of distributions is \textit{the spherical symmetry}, which means that the probability of $p'$ to fall on some point $p \in \R^{18}$, depends only on the distance of $p$ from the origin.

From the spherical symmetry of $p'$ and from the fact that $S^*$ is an intersection of halfspaces, whose boundaries pass through the origin, the probability
of $p'$ to appear in $S^*$ equals to the ratio between the hyper-surface area of $S^* \cap S^{17}$ and the hyper-surface area of the whole unit sphere $S^{17}$.

\paragraph{Dimension reduction from $\R^{18}$ to $\R^{4}$.} Finding the volume ratio in $\R^{18}$, as just defined, is a rather difficult task. Fortunately,
it can be significantly simplified by a dimension reduction. Let $N_1,\ldots,N_4$ be the normals of the hyperplanes defined by the system (2.1) so that
their positive direction points into $S^*$. $N_1,\ldots,N_4$ are linearly independent, so they can be a part of a basis of $\R^{18}$. Let
$V_5,\ldots,V_{18}$ be some vectors that complete $N_1,\ldots,N_4$ to a basis $B$ of $\R^{18}$ and are orthogonal to $N_1,\ldots,N_4$. By the basis
definition, each point $p \in \R^{18}$ can be uniquely represented as:
\[ 
p = \sum_{i=1}^{4} \alpha_i N_i + \sum_{i=5}^{18} \alpha_i V_i ,
\] for suitable real coefficients $\alpha_i$.

Note that a point $p$ belongs to $S^*$ iff its scalar products with $N_1,\ldots,N_4$ are all positive, therefore it can be determined whether $p \in S^*$ only from its coefficients $\alpha_1,\ldots,\alpha_4$. From that and the fact that $S^*$ is an intersection of halfspaces whose boundaries pass through the origin, we get that:
\[
\dfrac{Vol_{17}(S^*)}{Vol_{17}(S^{17})} = \dfrac{Vol_{3}(S^* \cap S^3)}{Vol_{3}(S^{3})},
\] where $S^3,S^{17}$ are the 3-dimensional and the 17-dimensional unit spheres, respectively, and $Vol_i(S)$ is the hyper-surface area of $S$, which may be the whole sphere $S^i$ or part of it.

In conclusion, the probability that $T = DT(P')$ equals (up to the small error noted earlier) to the ratio of the hyper-surface area of the part of the $S^3$ that belongs to the intersection of four halfspaces (whose boundaries pass through the origin and whose normals are linearly independent) that are determined by the system (2.1), and the whole $S^3$. An intersection of four such halfspaces with $S^3$ defines a \textit{spherical tetrahedron}. Thus our goal is to calculate the volume of a spherical tetrahedron in $R^4$ and divide it by the volume (hyper-surface area) of $S^3$.

\paragraph{Calculating the volume of a spherical tetrahedron in $\R^4$.} Surprisingly, until few years ago, no formula was known for calculating this volume. Only in 2010 Murakami \cite{M2012} gave such a formula (following several earlier works that handled tetrahedra of some special shape; see \cite{AM2014} for a survey of these results). The formula calculates the volume as a function of the six dihedral angles of the given spherical tetrahedron. In our case, these are the angles that are formed by all the pairs of the four supporting hyperplanes of the halfspaces that define $S^*$. For the convenience of the reader, Murakami's formula is presented in Appendix B of the thesis version \cite{KS2015}.

It is fairly easy to obtain the six dihedral angles of the spherical tetrahedron $S^* \cap S^3$ from $N_1,\ldots,N_4$. Concretely, the angle between the two hyperplanes with normals $N_i,N_j$ is $\pi - \arccos(N_i \cdot N_j)$, assuming that the normals are unit vectors.

\begin{figure}[h]

\centering

\includegraphics[width=0.95\textwidth]{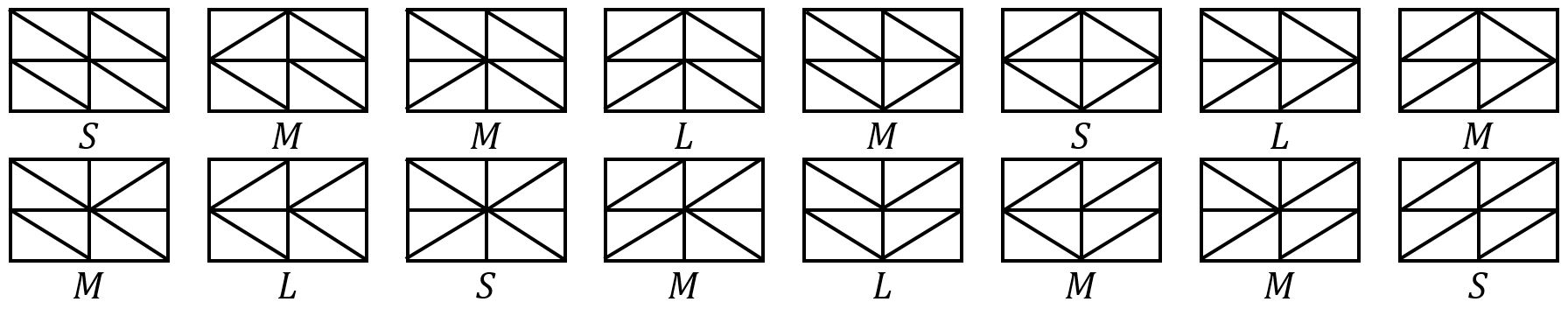}

\caption{\small The distribution of Delaunay triangulations of a grid $P$ with $m=2$. The letter below each triangulation represents its probability to appear as $DT(P')$, where $L=0.08422$, $M=0.06088$, and $S=0.04401$.}

\label{fig2.2}

\end{figure}

\paragraph{The results.} For each triangulation $T$ we calculated, by Murakami's formula, its probability to appear. The results are shown in Figure \ref{fig2.2}. As expected, these results match well the empirical findings from Appendix A.

In order to qualitatively describe the results, we say that two diagonals (or more) of $C(P)$ cells are identical iff they all have the same slope (+1 or -1), otherwise we say that they are opposite. By our results, the triangulations fall into three disjoint groups, according to their probability to appear as $DT(P')$. All the triangulations of each group share a common and unique property, as follows: Each triangulation that belongs to the group with the highest probability $(L = 0.08422)$, is such that each of its grid columns (or rows) contains identical diagonals, and adjacent columns (or rows) contain opposite diagonals. In contrast, each triangulation that belongs to the group with lowest probability $(S = 0.04401)$, is such that its cells that are located on the same grid diagonal (primary or secondary), contain identical diagonals. And for the last group with the middle probability $(M = 0.06088)$, all its triangulations have exactly three out of four identical diagonals.

\paragraph{\boldmath$m=3,4$.} For Delaunay triangulations of grids with $m > 2$, we were unable to calculate their probabilities explicitly. One reason for this is that we do not know of any explicit formula for calculating the volume of a spherical simplex on a sphere of dimension greater than 3. Empirical findings for these cases are presented in Appendix A.

From these findings we can see that the distribution of $DT(P')$ is clearly not uniform. Already for $m = 2$, the most common triangulations appear nearly
twice as many times as the rarest ones, and with the increase in $m$ this gap only grows. In addition, for different values of $m$ there are consistent
features in the common triangulations and other consistent features in the rare ones, similar to what has been observed for $m = 2$. These features are
discussed in the thesis \cite{KS2015}.

It would be interesting to try to extend these properties for larger grids.

\subsection{Large Grids (\boldmath$m \to \infty$)}

\begin{figure}[h]

\centering

\includegraphics[width=0.8\textwidth]{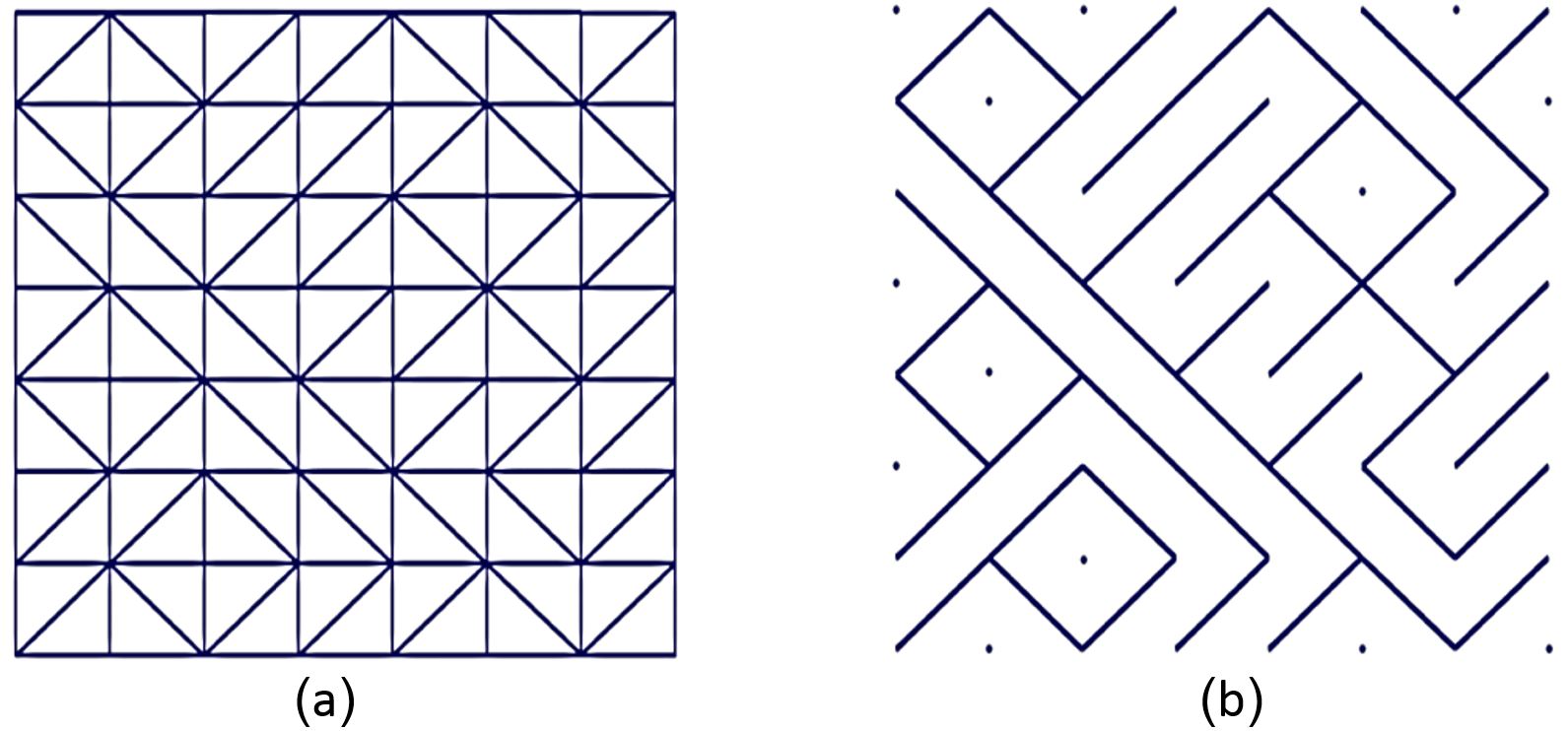}

\caption{\small (a) Triangulation $T$ of a grid $P$. (b) The graph $\widehat{T}$, that is obtained from $T$, with 19 connected components (including isolated vertices).}

\label{fig2.5}

\end{figure}

\begin{figure}[h]

\centering

\includegraphics[width=0.45\textwidth]{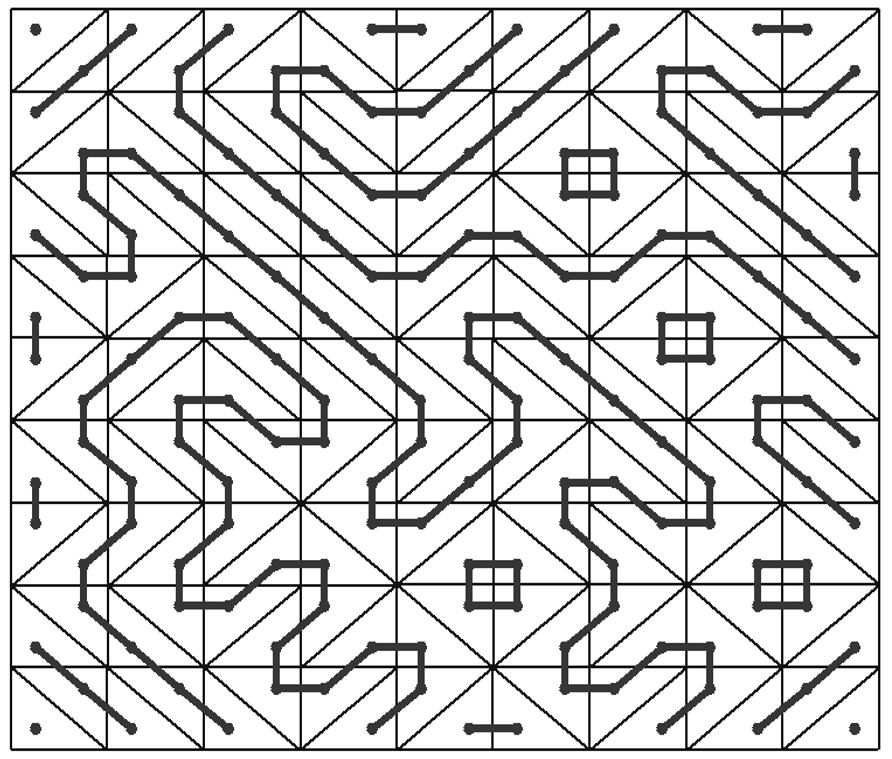}

\caption{\small A graph $G(T)$ (the thick lines) that is embedded in a triangulation $T$ (the thin lines).}

\label{fig2.6}

\end{figure}

\noindent We use some additional notations and definitions:

\begin{itemize}

\item $UT(P)$ -- The uniform random distribution of triangulations of $P$ in which, as for $DT(P')$, each grid cell is divided by a diagonal, this time with equal probability and independently.

\item $T$ -- A triangulation of $P$, of either type $DT(P')$ or $UT(P)$.

\item $\widehat{DT}(P'),\, \widehat{UT}(P),\, \widehat{T}$ -- subgraphs of $DT(P'),\, UT(P)$ and $T$, respectively, on all the grid vertices, such that their only edges are the diagonals of the cells of $C(P)$ (without the grid edges themselves). See Figure \ref{fig2.5}.

\item $G(T)$ -- A graph that is embedded in a triangulation $T$, either $DT(P')$ or $UT(P)$. Its vertices are the centers of the triangles of $T$, and two vertices are connected by an edge iff their corresponding triangles share a common horizontal or vertical edge, as depicted in Figure \ref{fig2.6}. Intuitively, $G(T)$ encodes adjacency between triangles of $T$ that are not through a common diagonal; the diagonals are ``walls'' that exclude adjacency.
(Informally, the edges of $G(T)$ form paths that traverse the ``maze'' formed by $\widehat{T}$.)

\item $CC(G)$ -- The number of the connected components of a graph $G$.

\end{itemize}

Because of the exponential dependence of the number of different triangulations on $m$, we cannot calculate the distribution of $DT(P')$ as before, even
empirically. Therefore, for large grids we take a different approach. Now our goal is just to show that in expectation $CC(\widehat{DT}(P')) <
CC(\widehat{UT}(P))$, thereby demonstrating that the two distributions are different. Empirical findings that substantiate this phenomenon are presented in
Table \ref{tbl2.1}. 

\begin{table}[h]

\centering

\includegraphics[width=0.6\textwidth]{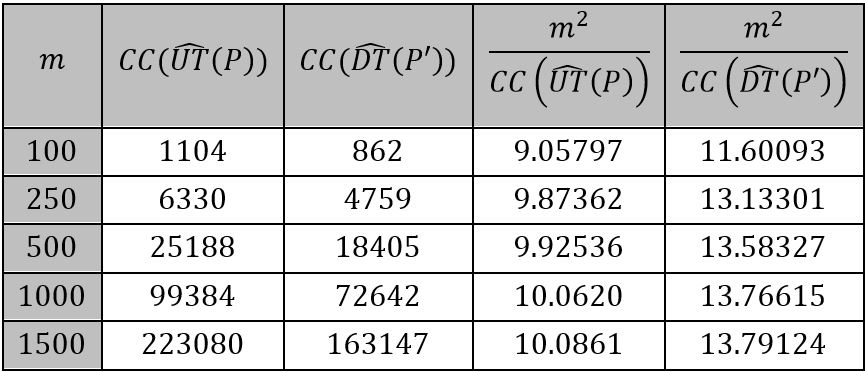}

\caption{\small The number of connected components and an estimate of the average size of a component of the graphs $\widehat{DT}(P'),\, \widehat{UT}(P)$,
that are obtained from a grid $P$ of size $(m+1) \times (m+1)$.}
\label{tbl2.1}
\end{table}

Observations about $G(T)$: it is a planar graph by definition, and the degrees of all of its vertices are at most 2. That is, $G(T)$ is a vertex-disjoint union of cycles and paths, where the paths start and end on the grid boundary. By a simple induction on the number of the cells of $C(P)$, it can be shown that $CC(\widehat{T}) = CC(G(T)) +
1$. Hence, it is enough to demonstrate that, in expectation, a typical cycle or path of $G(DT(P'))$ is longer than that of $G(UT(P))$.

\begin{figure}[h]

\centering

\includegraphics[width=0.22\textwidth]{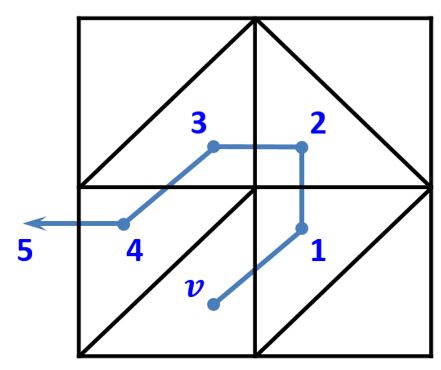}

\caption{\small A random walk of a particle between the cells of $C(P)$.}

\label{fig2.7}

\end{figure}

We substantiate this claim by thinking of a typical cycle of $G(T)$ as a \textit{random walk} of a particle between the cells of $C(P)$. In each discrete time $t \geq 0$
the particle is located in a cell $C_t \in C(P)$, in one of the two right-angled triangles $T_t$ that are obtained from $C_t$ by passing 
the diagonal in $T$. See Figure \ref{fig2.7}. The walk begins at an arbitrary cell $C_0 \in C(P)$, within one of its triangles $T_0$, and ends at the first time the particle returns back to $T_0$.

In order to estimate the expected length of these walks (up to the first return to the starting point), we approximated them using a \textit{Markov chain}
model, where each step is influenced only by the last four steps (this is an arbitrary choice, and is one of the reasons that the obtained results are only an approximation; informally, though, we expect that the effect of the diagonal in a cell $C_0$ on the choices in other cells rapidly decreases in $DT(P')$ as we wander away from $C_0$). For this purpose, for each triangulation type, we calculate a transition matrix of size $5 \times 5$ (since in four steps the particle cannot pass more than two cells in any direction). For $UT(P)$, because all of its probabilities are identical and independent, the matrix calculation is trivial. For $DT(P')$, we calculate it empirically using a computer. See the full version \cite{KS2015} for more details.

The approximations for the expectation and distributions of the length of a cycle in $G(T)$, restricting it to be at most 40, are summarized in Table \ref{tbl2.2}, where $T \in \{DT(P'),UT(P)\}$. These findings were obtained by a computer in two different ways:

\begin{enumerate}

\item By Markov chains, as briefly explained above.

\item By $10^5$ repeated iterations of a computer program. In each iteration the graphs $G(DT(P'))$, $G(UT(P))$ were calculated, where $P$ is a grid of size $m = 41$ (since in 40 steps the particle can not pass more than 21 rows or columns in any direction).

\end{enumerate}

\begin{table}[h]

\centering

\includegraphics[width=0.65\textwidth]{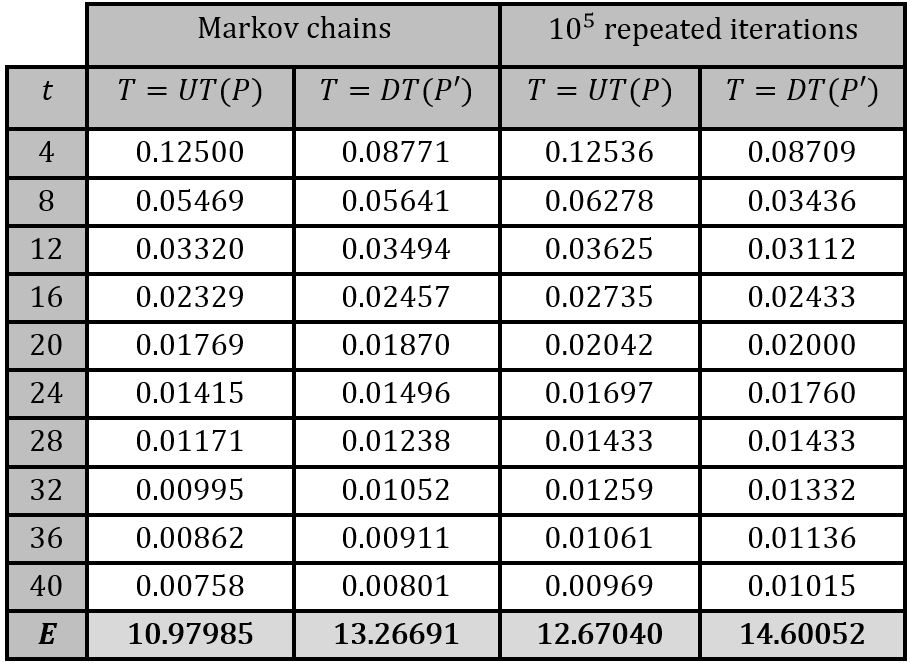}

\caption{\small Approximations for the expectation and distribution of the length $t$ of a cycle in $G(T)$ that passes through the central cell of $C(P)$,
restricting the length to be at most 40.}
\label{tbl2.2}
\end{table}

The reason for focusing only on cycles of a limited length (where the number 40 was chosen arbitrarily) is because otherwise many iterations might not stop
within a reasonable time. Following a known similar property of two-dimensional random walks, we believe, as explained in the full version \cite{KS2015}, that the expected length of a cycle, for both types of triangulations, tends to infinity.

Our empirical results show that the expected length of a cycle in $G(DT(P'))$ is larger than that of a cycle in $G(UT(P))$. In other words, up to the approximations that we have taken, it strongly suggests that $CC(\widehat{DT}(P')) < CC(\widehat{UT}(P))$. It should be noted that although the results, that were obtained by two different methods, are quite similar, there are, nevertheless, differences between them, some of which are quite significant. The reason for this lies in the built-in approximations of the Markov chains model.


\section{Regular Polygons}

\begin{figure}[h]

\centering

\includegraphics[width=0.3\textwidth]{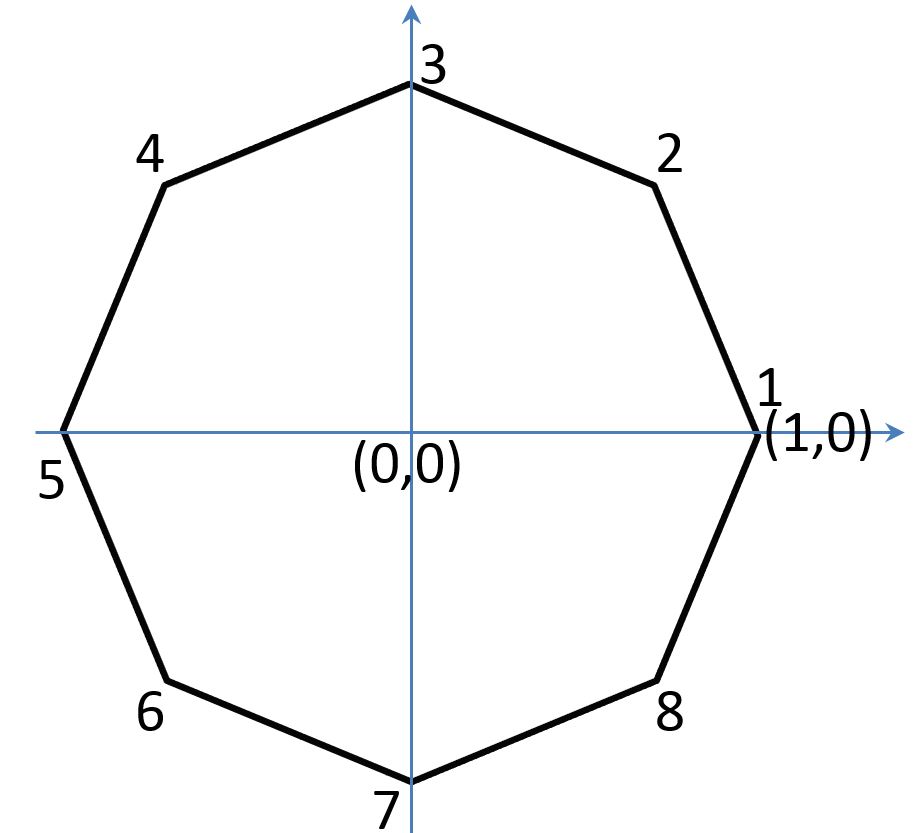}

\caption{\small A regular polygon $P$ with $n=8$ vertices.}

\label{fig3.1}

\end{figure}

In this chapter we take $P$ to be the set of vertices of a regular polygon, of size $n \geq 3$. For simplicity, we assume that the center of the polygon is at the origin and its rightmost vertex is at $(1,0)$. We enumerate the points of $P$ in counterclockwise (CCW) order, where the first point is the rightmost one, as illustrated in Figure \ref{fig3.1}.

Our main goal is to estimate the distribution of $DT(P')$. We managed to do it accurately and explicitly for $n \leq 7$, but for larger values we only have empirical results obtained via computer simulation. For some values of $n$, as an alternative, we tried to calculate the probability of a specific triangle to appear in $DT(P')$. Here too we only achieved partial (albeit nontrivial) success.

We emphasize that throughout our analysis, isomorphic triangulations, obtained by rotations and/or reflections of $P$, were treated as different triangulations.

\paragraph{The number of possible triangulations.} Note that from the empty circumcircle criterion, every triangulation of a regular polygon is a DT.
Naturally, we are interested in the number of different triangulations of a convex (and particularly regular) polygon, which, as is well known,
equals to $C_{n-2}$, the $(n-2)$-nd \textit{Catalan number} (see, e.g., \cite{SW}), where
\begin{align*} 
C_n = \dfrac{1}{n+1}\binom {2n}{n}.
\end{align*}


\subsection{\boldmath The distribution of $DT(P')$ for $3 \leq n \leq 16$}

\paragraph{\boldmath $n \in \{3,4,5\}$.} It is easy to see that all the possible triangulations are isomorphic to each other, so their probabilities to appear are equal. Using Catalan numbers, this common probability equals $\frac{1}{C_{n-2}}$, i.e., it is $1$, $\frac{1}{2}$, and $\frac{1}{5}$, respectively.

\paragraph{\boldmath $n \in \{6,7\}$.} Given a triangulation $T$ of $P$, we look for the probability that $T = DT(P')$. We present a general scheme to find $(n-3)$ halfspaces in $\R^{n-3}$ (as a sub-space of $\R^{2n}$ that represents the coordinate shifts of the $n$ points of $P$), such that their intersection with the unit sphere $S^{n-4}$ defines a spherical triangle/tetrahedron/simplex $S^*(T)$, such that the ratio between the volumes of $S^*(T)$ and $S^{n-4}$ is equal to the desired probability that $T = DT(P')$.

\begin{figure}[h]

\centering

\includegraphics[width=0.55\textwidth]{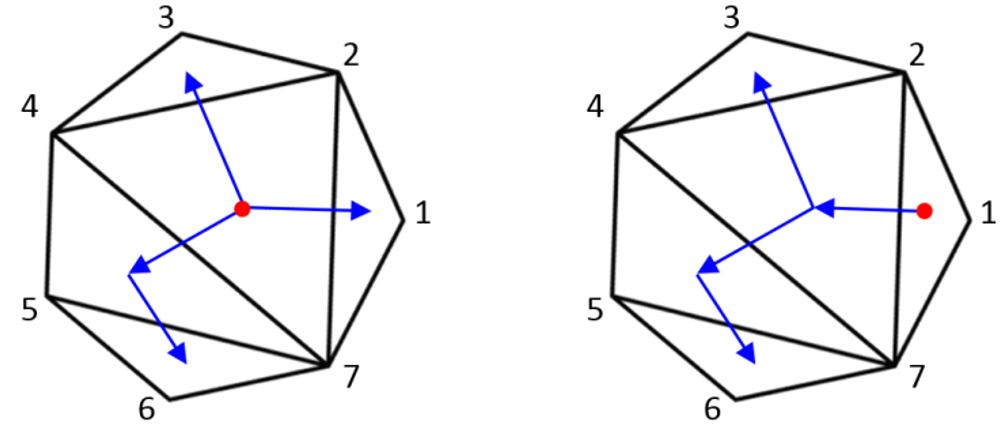}

\caption{\small Examples of possible match between a triangulation $T$ to a tree $tree(T)$. In each heptagon, the bold point inside a triangle (that is chosen arbitrarily) indicates the root of the tree. The arrows depict the ``branches" of the tree, i.e. the relation parent-child between the nodes of the tree that contain the matching triangles.}

\label{fig3.4}

\end{figure}

We begin by constructing a tree $tree(T)$ whose nodes are the triangles of $T$, as follows: First we fix an arbitrary triangle $\Delta_0$ of $T$ to be the
root of $tree(T)$. Next we construct the tree recursively top-down, i.e., from the root towards the leaves. Let $v$ be a node of $tree(T)$ that has already
been matched to a triangle $\Delta$ of $T$. Each triangle of $T$ that shares with $\Delta$ an edge and has not been yet matched to a node in $tree(T)$, is
taken to be (associated with) a child of $v$. See examples in Figure \ref{fig3.4}. From the fact that $T$ is a triangulation of a convex polygon, it follows that 
the resulting $tree(T)$ is indeed a tree. Let $\Delta_u$ denote the triangle of $T$ that was matched to the node $u$ in $tree(T)$.

We now show how to obtain from $tree(T)$ $n-3$ Incircle tests that force $T$ to be $DT(P')$. Every pair of nodes $u,v$ in $tree(T)$, such that $u$ is the parent of $v$, enforce the Incircle test that ensures that in $P'$ the vertex of $\Delta_v$ that is not a vertex of $\Delta_u$, is outside the circumcircle of $\Delta_u$. Since every triangulation of a regular $n$-gon has exactly $n-2$ triangles, we get that $tree(T)$ has exactly $n-3$ edges. Thus, in this way, we obtain $n-3$ different Incircle tests.

We claim that these $(n-3)$ tests indeed enforce $T$ to be $DT(P')$. If $T$ is $DT(P')$, then all of these tests certainly hold due to the empty circumcircle criterion of DT. 
Conversely, in order to have $T = DT(P')$, the circumcircle of every triangle $\Delta$ of $T$ should be empty of the points of $P'$ that are not the vertices of $\Delta$. Let $\Delta' \neq \Delta$ be another triangle of $T$ and $\delta'$ be a vertex of $\Delta'$ that is not a vertex of $\Delta$.
If $\Delta$ and $\Delta'$ share a common side, then $\delta'$ is outside the
circumcircle of $\Delta$, by one of the Incircle tests that $tree(T)$ defines. Otherwise, we focus on the path in $tree(T)$ that connects the nodes that
were matched to $\Delta$, $\Delta'$, and denote it as $\Delta'=\Delta_1, \Delta_2,\ldots, \Delta_k=\Delta$. For every $i$ the triangles $\Delta_i, \Delta_{i+1}$ share a common side and the circumcircle $C_{i+1}$ of $\Delta_{i+1}$ does not contain the vertex of $\Delta_i$ that is not a vertex of $\Delta_{i+1}$. It is easy to see that during the transition from $C_i$ to $C_{i+1}$, the part of $C_i$ that is at the side of $\Delta'$ shrinks and the other part expands. Thus, the vertex of $\Delta'$ that is not a vertex of $\Delta_2$, must be outside all of those circumcircles, including that of $\Delta$. We get that all the points of $P'$, except the vertices of $\Delta$, are outside its circumcircle, therefore $\Delta$ is a Delaunay triangle.

\begin{figure}[h]

\centering

\includegraphics[width=0.75\textwidth]{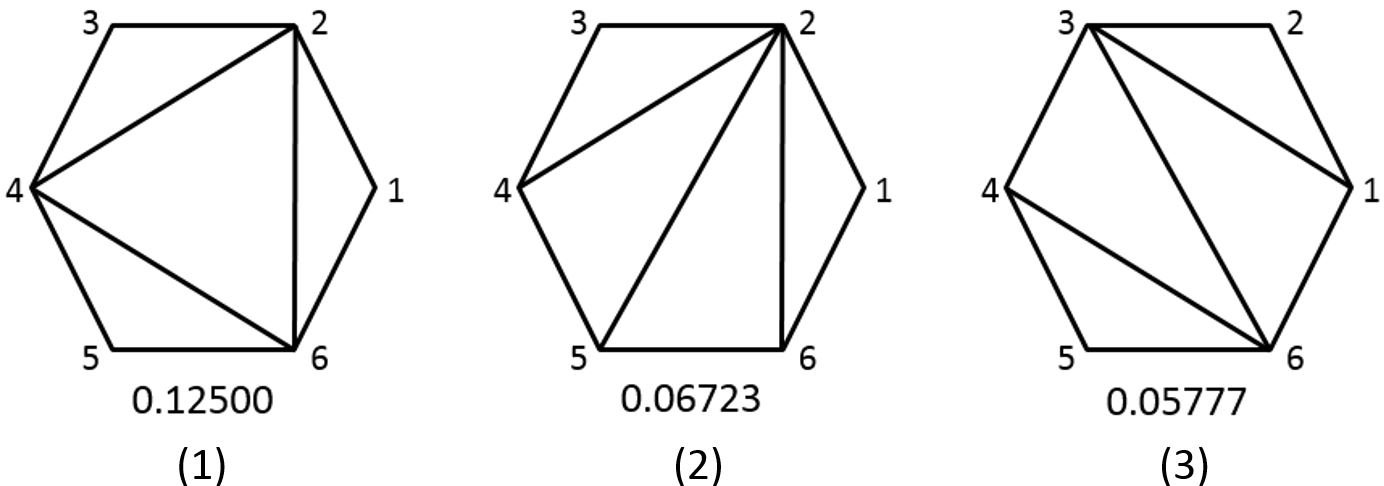}

\caption{\small Delaunay triangulations of a regular hexagon $P$. The number below each triangulation $T$ is the probability that $T=DT(P')$.}

\label{fig3.2}

\end{figure}

\begin{figure}[h]

\centering

\includegraphics[width=0.95\textwidth]{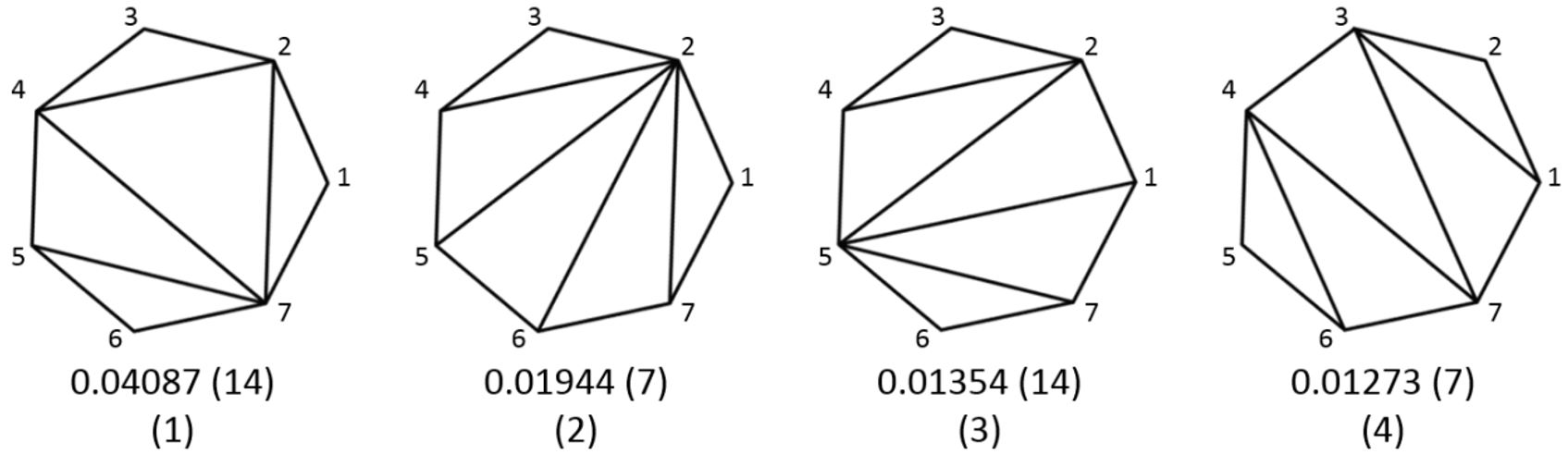}

\caption{\small Delaunay triangulations of a regular heptagon $P$. The number below each triangulation $T$ is the probability that $T=DT(P')$. On the right in parenthesis, is the number of the triangulations of $P$ that are isomorphic to $T$.}

\label{fig3.3}

\end{figure}

In order to get the desired probabilities, we applied the general scheme described above to hexagons and heptagons, for each triangulation $T$ that is
depicted in Figures \ref{fig3.2}, \ref{fig3.3} (every triangulation of a regular hexagon or heptagon is isomorphic to one of them). Then, for each $T$ we
calculated the surface area/volume of the corresponding spherical triangle/tetrahedron $S^*(T)$, as we described above for the uniform grid with $m = 2$
(using Murakami's formula for heptagons, and a simple formula for the area of spherical triangles, for hexagons). 
Finally, by dividing the surface area/volume of $S^*(T)$ by the surface area/volume of $S^{n-4}$, we get the desired probabilities.\footnote{We remind the reader that, as in the case of the grid, this interpretation of the probability is only a (good) approximation, because it ignores the effect of quadratic, and higher-order terms, on the sign of the determinant in the Incircle tests.}
 The results are depicted in Figures \ref{fig3.2} and \ref{fig3.3}. For
comparison, we note that the probabilities of a triangulation of a hexagon or a heptagon that is selected uniformly at random from all the 
respective 14 and 42 possible ones, are $1/14 \approx 0.0714$ and $1/42 \approx 0.02380$.

Here too, in order to verify the results, we performed empirical experiments. The findings are presented in Appendix A. As expected there is a strong
correlation between the theoretical and the empirical results, with one exception. We assume that the main reason for this exception is inaccuracies in the calculation of Murakami's formula.

The method described above can be applied to any $n \geq 8$ too, except that we do not know how to compute explicitly the volume of the resulting spherical simplex in $S^{n-4}$.

\paragraph{\boldmath $8 \leq n \leq 16$.} Here we only have empirical estimations that are based on computer simulations (see Appendix A). As just
mentioned, if we had an explicit formula for calculating the volume of a spherical simplex in spheres of dimension higher than 3, we would be able to expand our
theoretical analysis, given above for hexagons and heptagons, for polygons of size $n \ge 8$.

From our empirical findings it can be seen that the distribution of the triangulations $DT(P')$ is clearly not uniform. For example, for $n = 8$, the probability of any of the most common triangulations to appear is more than 10 times higher than that of any of the rarest ones, and more than 3.5 times higher than that of an average one. These gaps only expand with the increase in $n$.

Moreover, for different values of $n$ there are consistent features among the common and among the rare triangulations. A rare triangulation is typically
characterized by a zigzag pattern inside the polygon that is formed by its edges. For a typical common triangulation, if we associate with each
triangulation a list of the areas of its triangles, sorted in descending order, its list would be among the first ones in the lexicographical order. (Informally, "fat" triangles are more popular in $DT(P')$.)

\subsection{\boldmath The probability of a triangle to appear in $DT(P')$}

\paragraph{\boldmath $n \in \{6,7\}$.} Let $1 \leq i < j < k \leq n$ be vertices of $P'$. We now calculate the probability of the triangle $\Delta ijk$ to
appear in $DT(P')$. The condition for this is that for each point $q \in P' \setminus \{i,j,k\}$ it holds that $Incircle(i,j,k,q) < 0$. As mentioned
earlier, this is equal to the probability of a point to belong to the intersection of $n-3$ corresponding halfspaces in $\R^{n-3}$ (as a subspace of
$\R^{2n}$),
where the coordinates of the point are derived from the random shifts of the corresponding point of $P'$.
For $n \in \{6,7\}$ we get, for each $\Delta {ijk}$, three or four halfspaces, and again, by calculating the area/volume of the corresponding spherical triangle/tetrahedron, we obtain its probability.

\begin{figure}[h]
\centering
\includegraphics[width=0.8\textwidth]{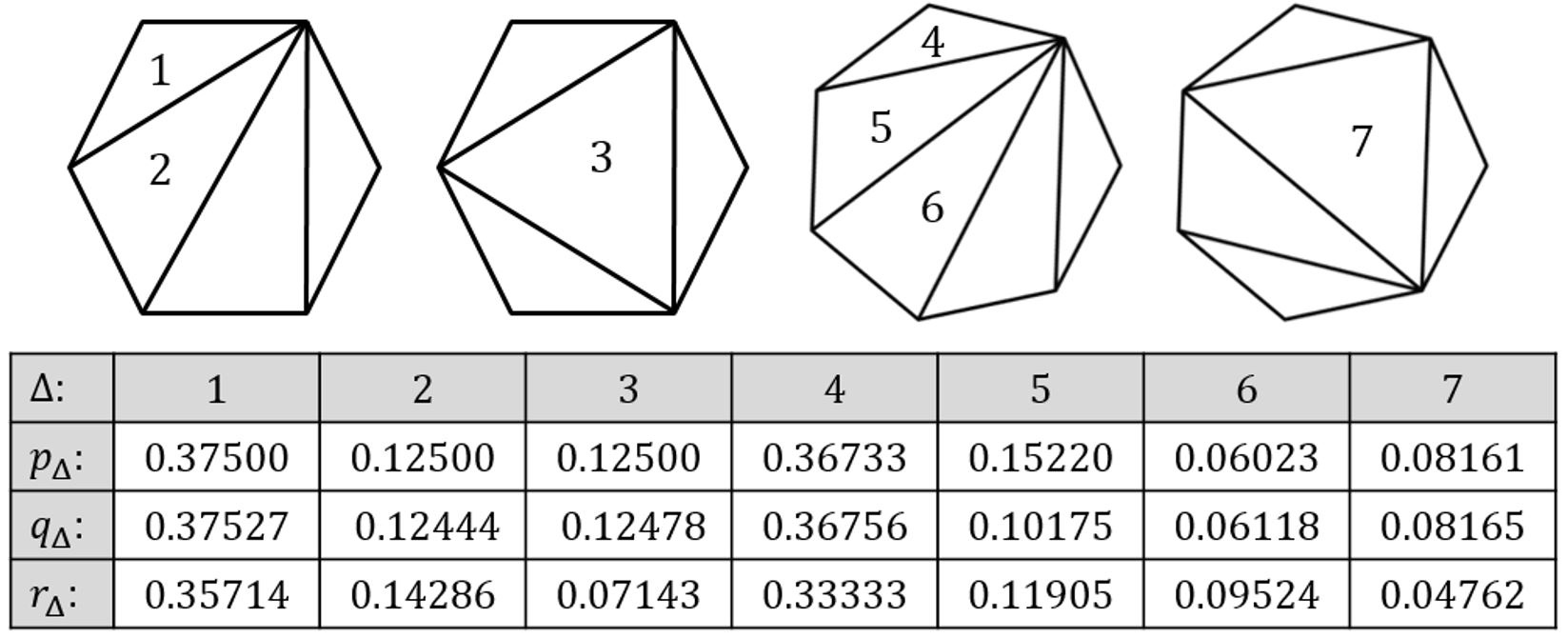}
\caption{\small The numbers in the table are the probabilities of the triangles $1 \leq \Delta \leq 7$, from the figure above, to appear in a
random triangulation of a regular hexagon/heptagon. $p_\Delta, q_\Delta$ are the probabilities of $\Delta$ to appear in $DT(P')$. $p_\Delta$ was calculated using Murakami's formula, as described in the text, and $q_\Delta$ was calculated empirically by $10^6$ iterations of a computer program. $r_\Delta$ is the probability of $\Delta$ to appear in a triangulation that is drawn uniformly at random, calculated using Catalan numbers.}
\label{fig3.5}
\end{figure}

The results are depicted in Figure \ref{fig3.5}. By symmetry, the probability of each $\Delta ijk$ equals to that of one of the triangles $1,\ldots,7$ in
the figure. For comparison, we added to the figure our empirical findings, and the probabilities of obtaining these triangles in a 
triangulation that is drawn uniformly at random, where the latter probabilities are easy to calculate using Catalan numbers.
(For example, the probability for triangle $7$ is $C_2/C_5$, and that of triangle $6$ is $C_2^2/C_5$.)

According to Figure \ref{fig3.5}, there is a strong correlation between the theoretical and empirical results. The only case where there is a relatively
large difference is in triangle 5. We did not analyse this in depth, but apparently, the reason for this are precision errors in the calculation of Murakami's formula.

In addition, as the table shows, all the triangles are distributed differently in $DT(P')$ than in the triangulation that is drawn uniformly at random. One
may roughly say that for Delaunay triangulations, the probabilities of the ``cornermost'' triangles (triangles 1 and 4) and the ``biggest'' ones (triangles 3 and 7) are greater than for the triangulation that is drawn uniformly at random, while for the rest of the triangles the opposite is true.

\begin{figure}[h]

\centering

\includegraphics[width=0.4\textwidth]{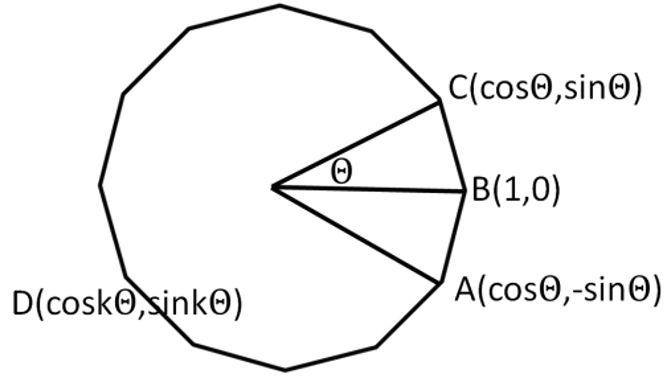}

\caption{\small A regular polygon $P$ with many vertices.}

\label{fig3.6}

\end{figure}

\paragraph{\boldmath $n \to \infty$.} Let $A, B, C \in P$ be three different vertices that are arranged in CCW order, as depicted in Figure \ref{fig3.6} (the figure only depicts the case where $A, B, C$ are consecutive). We want to calculate the probability of $\Delta ABC$ to appear in $DT(P')$. By the empty circumcircle criterion, this happens iff after the perturbation each point $D \in P' \setminus \{A,B,C\}$ is located outside the circumcircle of $\Delta ABC$, i.e., all the $n - 3$ corresponding
Incircle tests are negative simultaneously.

Let us express each point $Q \in P'$ by $(\cos{k_Q\theta} + x_Q, \sin{k_Q\theta} + y_Q)$, where $0 \leq k_Q \leq |P| - 1$, $\theta = \frac{2\pi}{|P|}$, and $x_Q, y_Q$ are the shifts in the $x$- and $y$-coordinates of point $Q$, as a result of the normal perturbation. (As usual, all the normal variables $x_Q, y_Q$ are independent and identically distributed.) We are looking for the probability of all the following $n - 3$ inequalities to hold simultaneously, for $2 \leq k_D \leq n - 2$:
$$
\begin{vmatrix}
\cos{k_A\theta + x_A} & \sin{k_A\theta} + y_A & (\cos{k_A\theta + x_A})^2 + (\sin{k_A\theta} + y_A)^2 & 1\\
\cos{k_B\theta + x_B} & \sin{k_B\theta} + y_B & (\cos{k_B\theta + x_B})^2 + (\sin{k_B\theta} + y_B)^2 & 1\\
\cos{k_C\theta + x_C} & \sin{k_C\theta} + y_C & (\cos{k_C\theta + x_C})^2 + (\sin{k_C\theta} + y_C)^2 & 1\\
\cos{k_D\theta + x_D} & \sin{k_D\theta} + y_D & (\cos{k_D\theta + x_D})^2 + (\sin{k_D\theta} + y_D)^2 & 1\\
\end{vmatrix} < 0.
$$
After simplifying the third column of each determinant, using the identity: $\sin^2\theta + \cos^2\theta = 1$, and excluding terms of second or higher order, we get:
$$
\begin{vmatrix}
\cos{k_A\theta + x_A} & \sin{k_A\theta} + y_A & x_A\cos{k_A\theta} + y_A\sin{k_A\theta} & 1\\
\cos{k_B\theta + x_B} & \sin{k_B\theta} + y_B & x_B\cos{k_B\theta} + y_B\sin{k_B\theta} & 1\\
\cos{k_C\theta + x_C} & \sin{k_C\theta} + y_C & x_C\cos{k_C\theta} + y_C\sin{k_C\theta} & 1\\
\cos{k_D\theta + x_D} & \sin{k_D\theta} + y_D & x_D\cos{k_D\theta} + y_D\sin{k_D\theta} & 1\\
\end{vmatrix} < 0.
$$

Let $z_i = x_i\cos{k_i\theta} + y_i\sin{k_i\theta}$, for $i \in \{A,B,C,D\}$. From the spherical symmetry of the multivariate normal distribution, all the variables $z_i$ are drawn from the same distribution function as $x_i,y_i$, and they, as $x_i,y_i$, are independent. After evaluating all the determinants according to their third column, we get:
\begin{multline*}\tag{3.1}
z_A
\begin{vmatrix}
\cos{k_B\theta + x_B} & \sin{k_B\theta} + y_B & 1\\
\cos{k_C\theta + x_C} & \sin{k_C\theta} + y_C & 1\\
\cos{k_D\theta + x_D} & \sin{k_D\theta} + y_D & 1\\
\end{vmatrix}
- z_B
\begin{vmatrix}
\cos{k_A\theta + x_A} & \sin{k_A\theta} + y_A & 1\\
\cos{k_C\theta + x_C} & \sin{k_C\theta} + y_C & 1\\
\cos{k_D\theta + x_D} & \sin{k_D\theta} + y_D & 1\\
\end{vmatrix} + \\
+ z_C
\begin{vmatrix}
\cos{k_A\theta + x_A} & \sin{k_A\theta} + y_A & 1\\
\cos{k_B\theta + x_B} & \sin{k_B\theta} + y_B & 1\\
\cos{k_D\theta + x_D} & \sin{k_D\theta} + y_D & 1\\
\end{vmatrix}
- z_D
\begin{vmatrix}
\cos{k_A\theta + x_A} & \sin{k_A\theta} + y_A & 1\\
\cos{k_B\theta + x_B} & \sin{k_B\theta} + y_B & 1\\
\cos{k_C\theta + x_C} & \sin{k_C\theta} + y_C & 1\\
\end{vmatrix}
< 0.
\end{multline*}

Let $S(\Delta)$ denote the area of a triangle $\Delta$. After dividing each inequality by two, and again excluding terms of second or higher order, we get:
$$
z_D > \dfrac{1}{S(\Delta ABC)}\Big(z_AS(\Delta BCD) - z_BS(\Delta ACD) + z_CS(\Delta ABD)\Big).
$$
Due to the symmetry of the normal distribution, we get that the desired probability is equal to the integral:
\begin{multline*}\tag{3.2}
\iiint \bigg(\prod_{D \in P \setminus \{A,B,C\}} \Phi \Big(- \dfrac{z_AS(\Delta BCD) - z_BS(\Delta ACD) + z_CS(\Delta ABD)}{S(\Delta ABC)}\Big)\bigg) \cdot \\ \,\varphi(z_A)\,\varphi(z_B)\,\varphi(z_C)\,dz_A\,dz_B\,dz_C,
\end{multline*} where $\varphi(x)$ is the standard normal density function, and $\Phi(x)$ is the standard normal distribution function.\footnote{Technically, our shifts $x_i,y_i$, and thus $z_i$ too, are drawn from a scaled-down version of the normal distribution. Nevertheless, since the inequalities in (3.1) are linear and homogeneous, we may assume that they are drawn from the standard normal distribution.}

For the ``corner'' triangle $\Delta 123$, we approximately calculated the integral (3.2). The results are given in Table \ref{tbl3.1};
they show that the probability of the triangle to appear in $DT(P')$ is relatively high and nearly a constant for every tested value of $n$.

\begin{table}[h]
\centering
\includegraphics[width=0.8\textwidth]{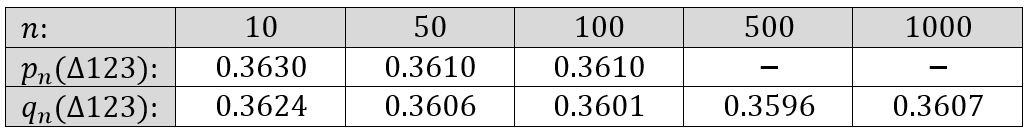}
\caption{\small $p_n(\Delta123), q_n(\Delta123)$ are the probabilities of the corner triangle $\Delta123$ to appear in $DT(P')$, where $n$ is the size of $P$. $p_n(\Delta123)$ are the results of numeric evaluations of the integral (3.2), where the values $n \in \{500,1000\}$ are too large for the numeric evaluation, and $q_n(\Delta123)$ were obtain by $10^6$ repeated iterations of a computer program.}
\label{tbl3.1}
\end{table}

Here is an informal attempt to explain this phenomenon.
For the corner triangle $\Delta ABC = \Delta 123$ and a specific $D \in P \setminus \{A, B, C\}$, we denote:
\[ 
\alpha = S(\Delta BCD), \quad  \beta = S(\Delta ABD), \quad \epsilon = S(\Delta ABC) ,
\] 
and note that $S(\Delta ACD) = \alpha + \gamma - \epsilon$. The argument of $\Phi(\cdot)$ for $D$ is
$
\big(
-z_B + \frac{\alpha}{\epsilon}(z_B - z_A) + \frac{\gamma}{\epsilon}(z_B - z_C)
\big)
$. We have $\epsilon = \Theta\left(\frac{1}{n^3}\right)$, and, for most of the points $D$, we have $\alpha = \Theta\left( \frac{1}{n}\right)$,\
and $\gamma = \Theta\left(\frac{1}{n}\right)$. So $\frac{\alpha}{\epsilon},\frac{\gamma}{\epsilon} = \Theta(n^2)$. 
Particularly, in the domain where $z_B > z_A,z_C$, an event that happens with probability $\frac{1}{3}$, a lot of values of 
$\Phi(\cdot)$ in the product integrand tend to 1, and this also happens in the other domains, with a non-negligible probability. These considerations give
an intuitive explanation of the values in Table \ref{tbl3.1}. (We note that, for significantly non-corner triangles, $S(\Delta ABC)$ is much larger, and therefore the domain in which the integral in (3.2) is close to $1$ is smaller, and thus the value of the integral gets smaller.)

It is interesting to compare this probability (for a corner triangle) to its probability to appear in a triangulation of $P$ that is drawn uniformly at random. Using Catalan numbers, one can show that the latter probability approaches $\frac{1}{4}$ for large $n$. In other words, the probability of a corner triangle to appear in a random Delaunay triangulation $DT(P')$ (in our model) is significantly greater than its probability to appear in any triangulation that is drawn uniformly at random.

We conclude by raising the open problem of proving rigorously that the limiting probability of $\Delta 123$ to appear in $DT(P')$, as $n\to\infty$, exists,
and to find an explicit expression for it.


\section*{Appendix A: Empirical Findings}

All the results in this appendix were obtained by many repeated iterations of computer programs, using a pseudo-random generator. For different point sets
$P$ in the plane, for each $T \in DT(P)$, we estimated the probability that $DT(P') = T$, where $P'$ is a (random) set of points that were obtained from $P$
by a normal perturbation.

\subsection*{Grids}

Let $P$ be a uniform $(m + 1) \times (m + 1)$ grid of points in the plane, as described in Section 2. For readability, we naturally represent each $DT(P)$
as an $m \times m$ binary matrix, where each of the matrix cells corresponds to a different cell $c \in C(P)$. ``1'' (resp., ``0'')
in the matrix means that in $DT(P)$ the corresponding cell $c$ is divided by a diagonal with positive slope (`/') (resp., (`\textbackslash{}')).

\begin{figure}[h]

\centering

\caption*{{\boldmath $m = 2$}. The number below each matrix is the relative number of times the corresponding triangulation appeared out of $10^6$ iterations. These results can be compared to the accurate results from Section 2.}

\includegraphics[width=0.9\textwidth]{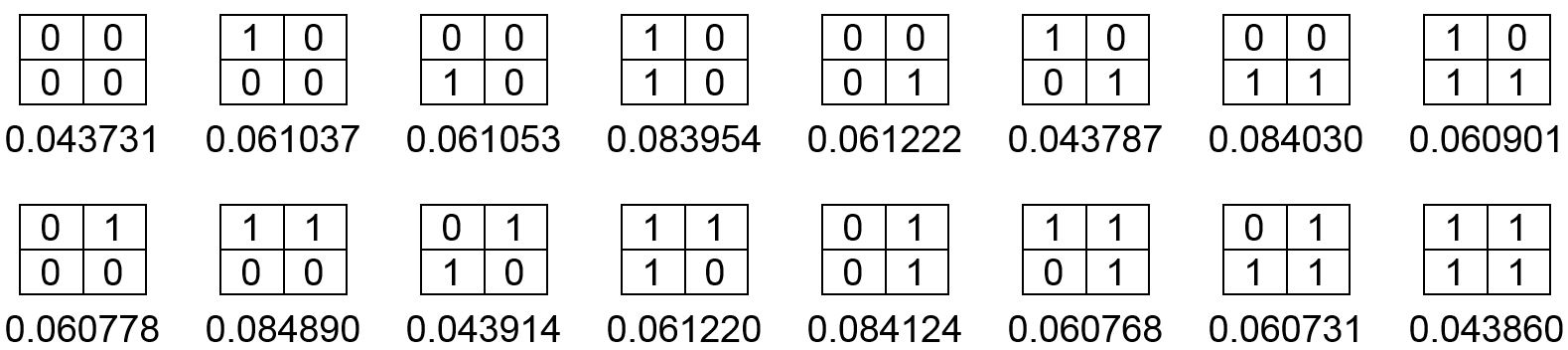}

\label{figA1.1}

\end{figure}

\clearpage

\begin{figure}[h]

\centering

\caption*{{\boldmath $m = 3$}. The table below depicts the 20 most common triangulations. The number below each matrix is the number of times the corresponding triangulation appeared out of $10^6$ iterations.}

\includegraphics[width=0.95\textwidth]{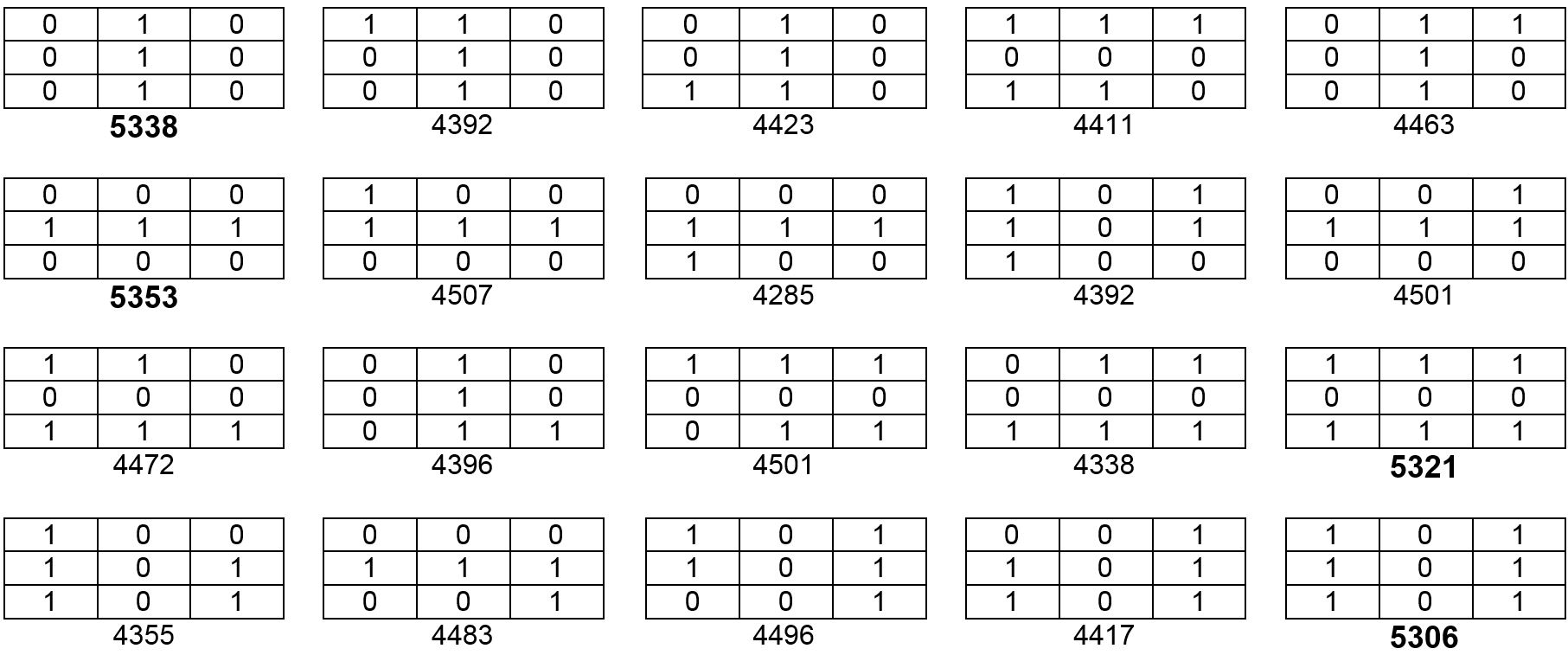}


\label{figA1.2}

\end{figure}

In principle, all these triangulations belong to two equivalence classes. In the first class, the most frequent one, the diagonals are equal in the two extreme rows or in the two extreme columns, and are opposite in the middle row or in the middle column. In the second class the structure is similar, except for a diagonal in one extreme cell, that changes its direction. \\

\begin{figure}[h!]

\centering

\caption*{{\boldmath $m = 3$}. The table below depicts the 20 rarest triangulations. The number below each matrix is the number of times the corresponding triangulation appeared out of $10^6$ iterations.}

\includegraphics[width=0.95\textwidth]{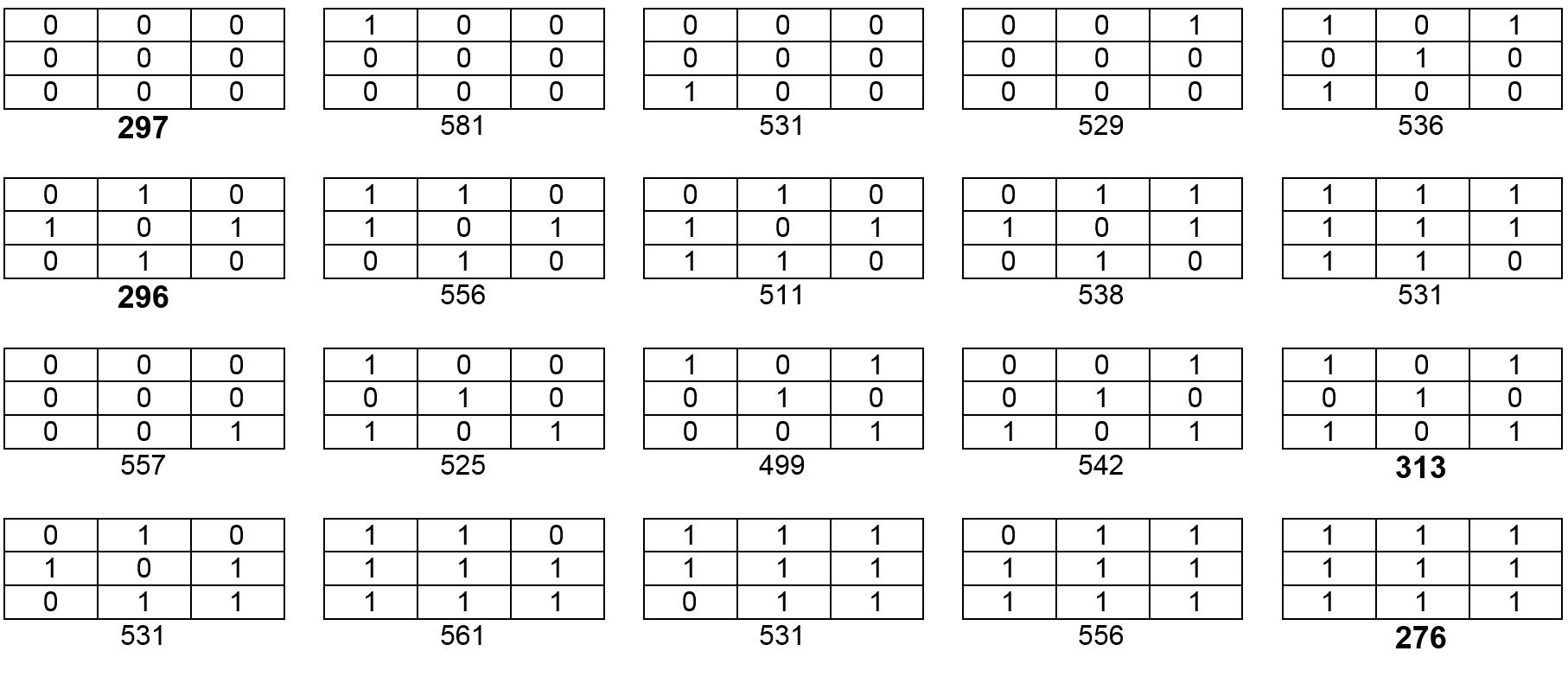}

\label{figA1.3}

\end{figure}

\clearpage

\begin{figure}[h]

\centering

\caption*{{\boldmath $m = 4$}. The table below depicts the 14 most common triangulations. The number below each matrix is the number of times the corresponding triangulation appeared out of $10^6$ iterations.}

\includegraphics[width=1\textwidth]{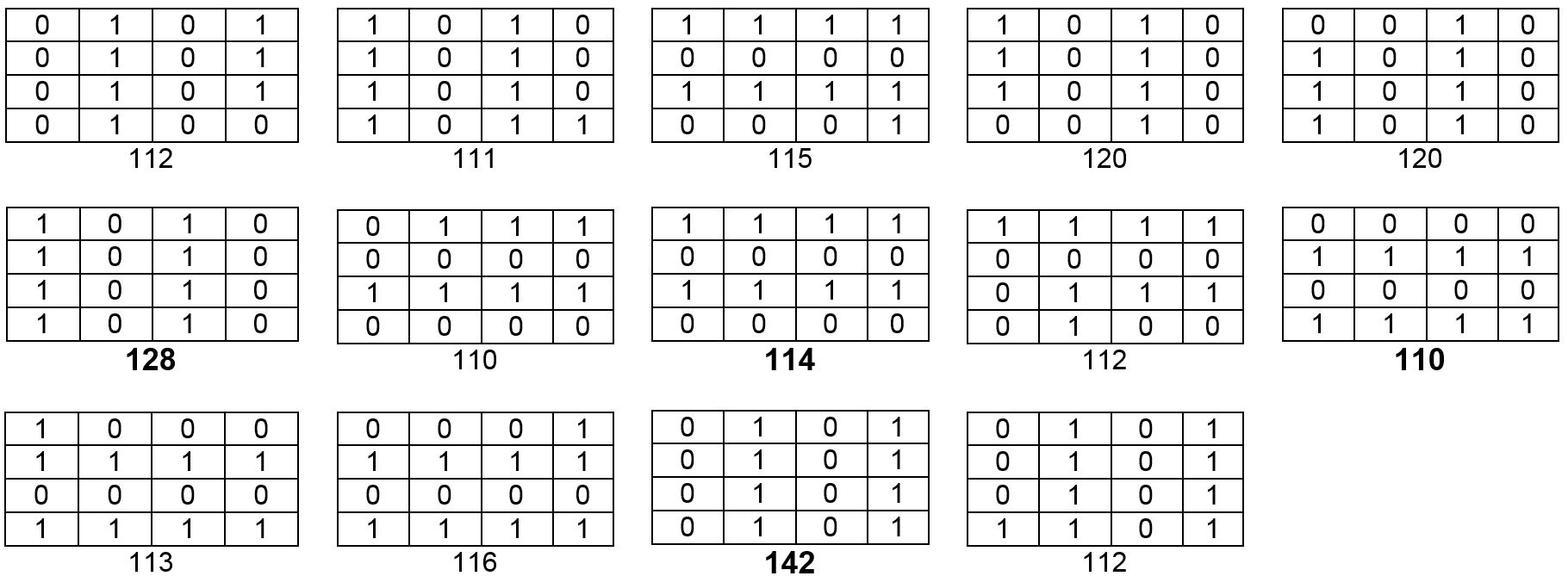}


\label{figA1.4}

\end{figure}

Here too the structure is similar to the case of $m = 3$, which consists (mostly) of rows (or columns) with identical diagonals, that are reversed between consecutive rows (or columns).




\subsection*{Regular Polygons}

Let $P$ be a regular polygon of size $n$, as defined in Section 3. The number below each triangulation $T$ is the number of times it appeared out of $10^6$ iterations. In parenthesis is the number of the triangulations that are isomorphic to $T$.

\begin{figure}[h]

\centering

\caption*{{\boldmath $n = 7$}. \hfill $ $}

\includegraphics[width=0.9\textwidth]{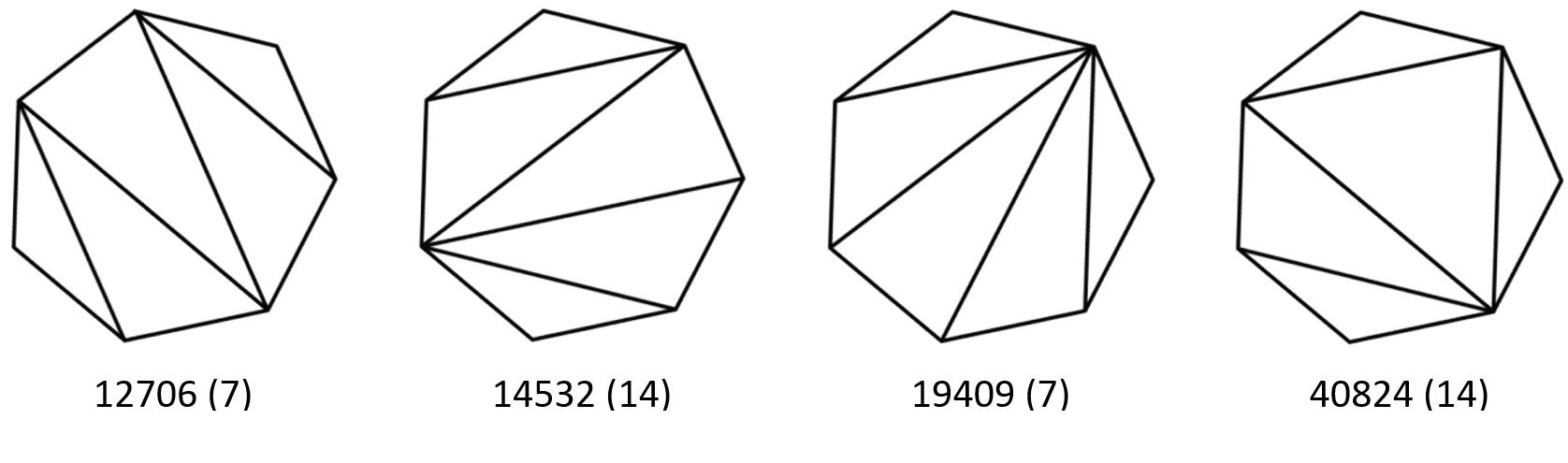}

\label{figA2.1}

\end{figure}

\clearpage

\begin{figure}[!h]

\centering

\caption*{{\boldmath $n = 8$}. \hfill $ $}

\includegraphics[width=0.95\textwidth]{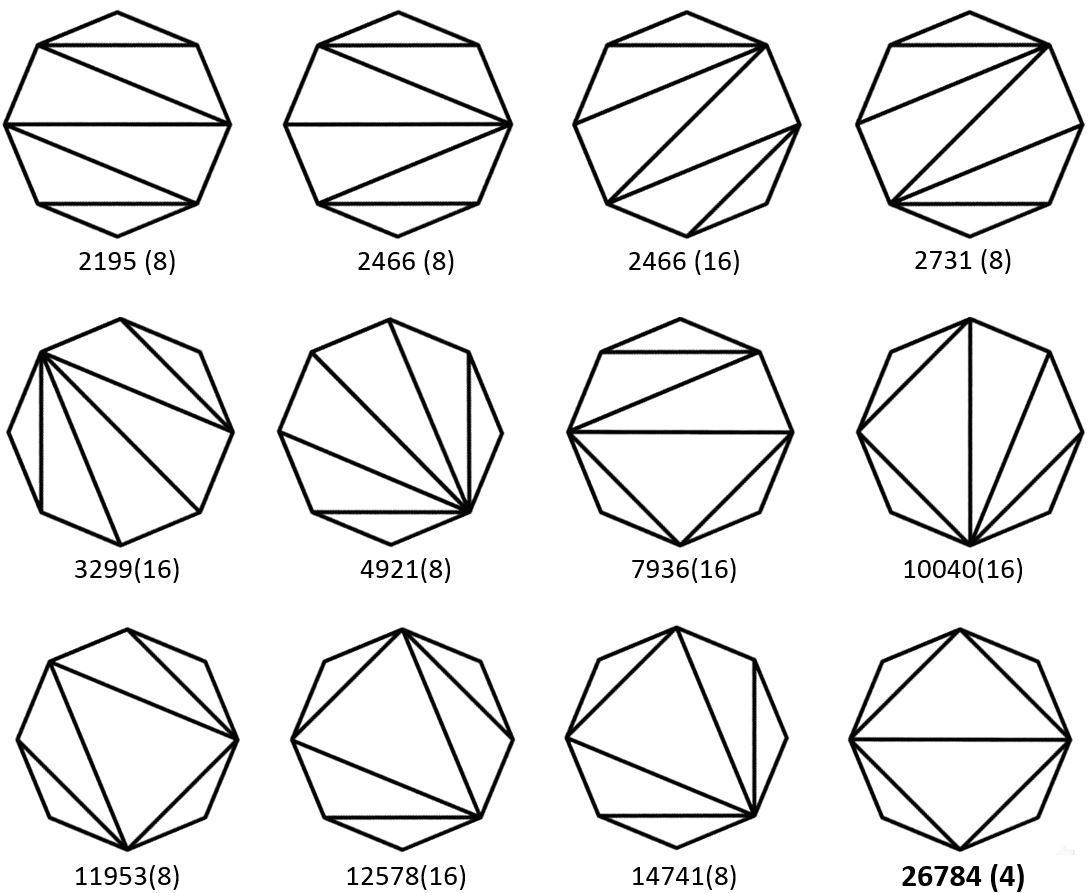}

\label{figA2.2}

\end{figure}


\clearpage

\begin{figure}[!h]

\centering

\caption*{{\boldmath $6 \leq n \leq 16$}. For each $n$, the most common triangulation is depicted. Below each triangulation $T$, $p$ is the relative number of times $T$ appeared out of $10^6$ iterations, and $q = \frac{1}{C_{n-2}}$ is the average probability of any triangulation to appear as $DT(P')$.}

\includegraphics[width=0.95\textwidth]{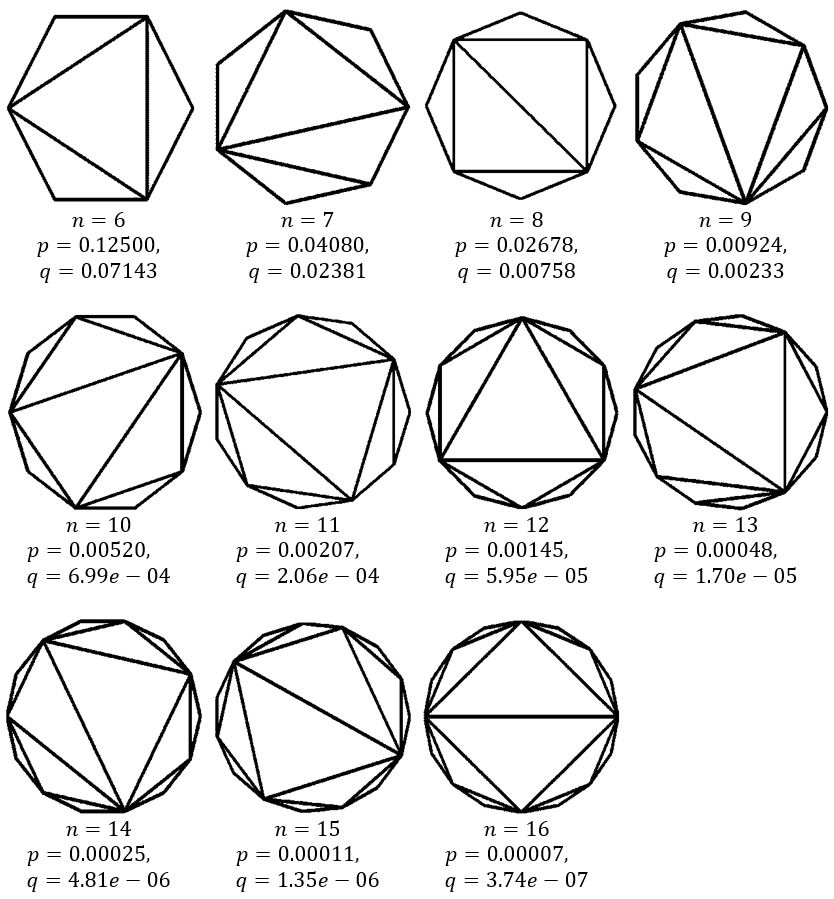}

\label{figA2.3}

\end{figure}


\end{document}